\documentclass[a4paper]{jpconf}

\usepackage{graphicx}
\usepackage{amssymb}
\usepackage{subfig}
\usepackage{cite}

\begin{document}

\title{Overview of electromagnetic probes in ultra-relativistic heavy-ion collisions}

\author{Jean-Fran\c{c}ois Paquet}

\address{Department of Physics and Astronomy, Stony Brook University, Stony Brook, New York 11794, USA}

\ead{jean-francois.paquet@stonybrook.edu}

\begin{abstract}
An introductory overview of electromagnetic probes in ultra-relativistic heavy-ion collisions is provided. Experimental evidence supporting the production of thermal photons and dileptons in heavy-ion collisions at the 
Relativistic Heavy Ion Collider (RHIC) and the Large Hadron Collider (LHC)
are reviewed. Thermal electromagnetic emission evaluated from hydrodynamical models of collisions is discussed.
\end{abstract}

\section{Introduction}

The production of a deconfined nuclear plasma in heavy-ion collisions currently under investigation at 
RHIC and the LHC
provides a window into the many-body properties of Quantum Chromodynamics. The exceedingly short lifetime of this deconfined plasma prevents the use of external probes to study its properties, which must rather be inferred from the imprint left by the plasma on particle flying out of the collision region.

The standard picture of the evolution of a heavy-ion collision begins with a relatively rapid approach to local equilibrium ($\tau \sim 0.1-1$~fm) for the mid-rapidity region of the collision. This locally quasi-equilibrated plasma undergoes a rapid hydrodynamical expansion until its density becomes too small to maintain local equilibrium ($\tau \sim 10$~fm). This last stage is thought to happen after the on-set of confinement, allowing the late evolution of the medium to be described in terms of hadronic degrees of freedom. While multiple uncertainties remain in this picture of the evolution of the deconfined plasma, in particular for the early stage of the collision, the overall success of hydrodynamical models of heavy-ion collisions in describing a wide array of soft hadron measurements suggests that the main features of this picture are credible\footnote{This work focuses on heavy-ion collisions at the LHC as well as the top RHIC energy (Au-Au at $\sqrt{s_{NN}}=200$~GeV), where evidence for the above picture of the evolution of collisions are most convincing at the moment.} (see e.g. Refs.~\cite{Denicol:2016nrw,deSouza:2015ena} for recent overviews of hydrodynamical modelling of heavy-ion collisions).

The deconfined plasma created at RHIC and the LHC does not generally exceed $\sim 10$~fm in size in the direction transverse to the collision axis. Such a transverse size is small compared to the mean-free-path of energetic photons and dileptons\footnote{
The equilibration time for photons and dileptons is of order~\cite{Kapusta:1991qp,Wong:1995jf}  $\tau_{eq}\sim \frac{f^{(0)}_B(k)/(2\pi)^3}{d^3 \Gamma/d k^3}$ where $f^{(0)}_B(k)$ is the Bose-Einstein distribution (which, up to a degeneracy factor, is the equilibrium distribution of photons and dileptons), and $d^3 \Gamma/d k^3$ is the thermal photon/dilepton emission rate. For soft photons of $1-3$~GeV of energies and typical $150-400$~MeV plasma temperatures, $\tau_{eq}\sim 100-500$~fm, about an order of magnitude larger than the lifetime of the plasma.
}. This leads to the conclusion that the deconfined plasma, while electrically charged, is optically thin for such probes. Electromagnetic probes produced at any point during the lifetime of the plasma can thus escape with minimal interaction. 

Photons and dileptons produced by the expanding deconfined medium are referred to as ``thermal'', since the plasma they are emitted from is close to local thermal equilibrium. The rate of electromagnetic probe production and the momentum distribution of these probes varies according to the local properties of the plasma --- the plasma's temperature for example, or the extent of its local deviation from equilibrium. A number of theoretical studies have shown the sensitivity of thermal electromagnetic probes to the properties of the deconfined nuclear matter produced in heavy-ion collisions (see Refs.~\cite{Shen:2015nto,Shen:2016odt} for a recent overview). The enhanced dependence of electromagnetic observables, compared to hadronic ones, on the earlier stage of the plasma evolution is of particular interest (see e.g. Ref.~\cite{Vujanovic:2016anq}). In consequence, experimental measurements of thermal electromagnetic probes can provide valuable information about the properties of hot deconfined nuclear matter. 

Yet this information is not always easily accessible, since thermal electromagnetic probes cannot be singled out experimentally from other sources of electromagnetic radiation, and these other sources often outshine the thermal signal. To better understand the different sources of photons and dileptons present in heavy-ion collisions, and how they compete with thermal probes in measurements, a brief overview of experimental electromagnetic observables is first provided. The production of thermal photons and dileptons from a hydrodynamical model is discussed next.

\section{Experimental observables and sources of electromagnetic probes}

Photon and dilepton production in relativistic \emph{proton-proton} (p-p) collisions is in general well understood theoretically~\cite{Aurenche:2006vj,McGaughey:1999mq}. One expects every source of electromagnetic radiation present in p-p collisions to also be produced in heavy-ion collisions, in addition to sources such as thermal photons and dileptons that result from the formation of a deconfined plasma. The term ``cold electromagnetic sources'' can be used to describe emissions already present in p-p collisions, while ``hot sources'' can be used for sources produced in connection with the hot plasma. This classification is not always clear-cut but is still useful.
In the simplest cases, specific cold electromagnetic sources in heavy-ion collisions are related to the same sources in proton-proton collisions by binary nucleon collision scaling: that is, for certain sources of radiation, a heavy-ion collision can be seen as an incoherent superposition of proton-proton collisions.
Relativistic proton-\emph{nucleus} collisions are used frequently as an intermediate step to calibrate ``cold'' sources of photons and dileptons from p-p collisions to heavy-ion ones (e.g. Ref.~\cite{Adare:2012vn}).

\paragraph{Dileptons}

\begin{figure}[tbp]
        \centering
				\includegraphics[width=0.32\textwidth]{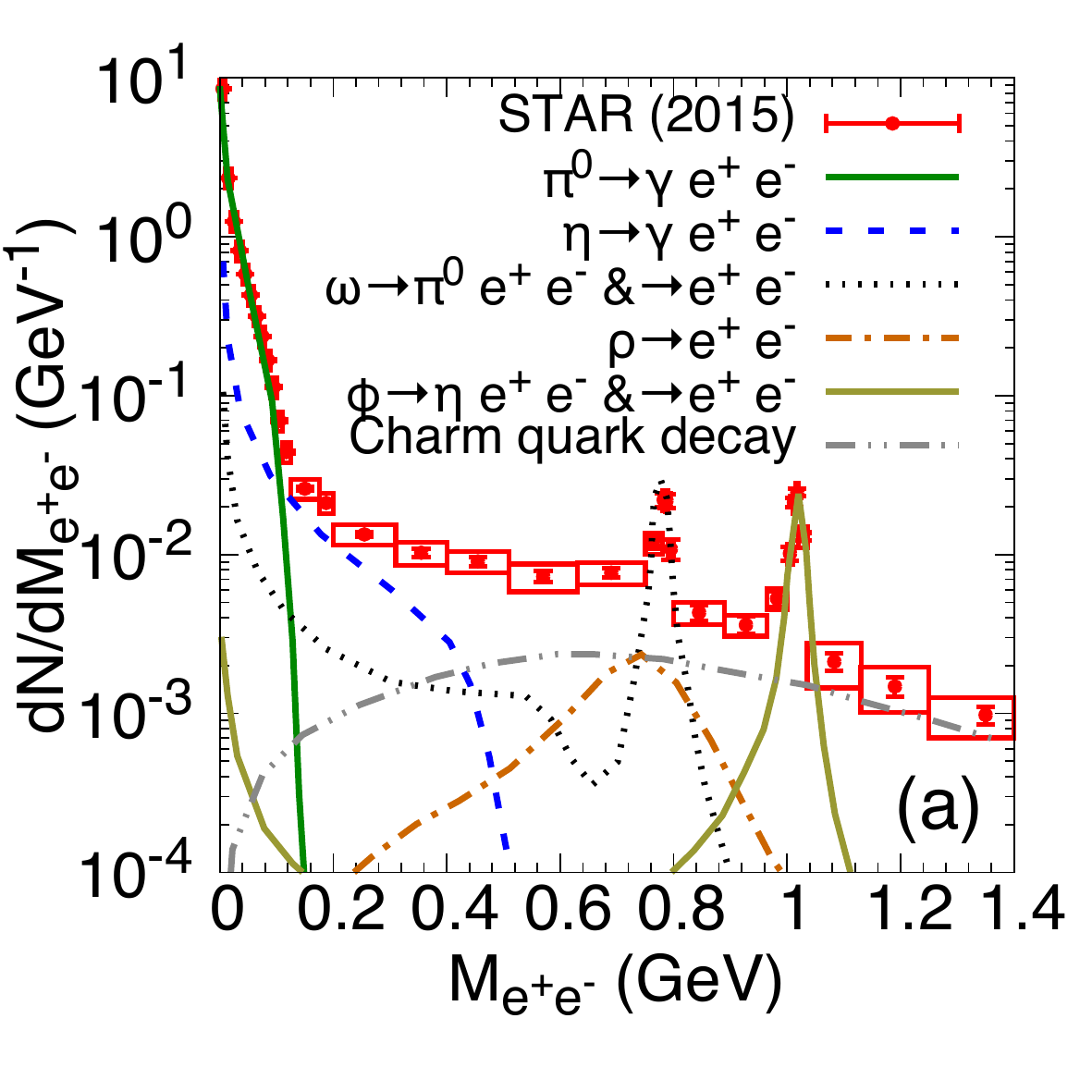}
				\hfill
				\includegraphics[width=0.32\textwidth]{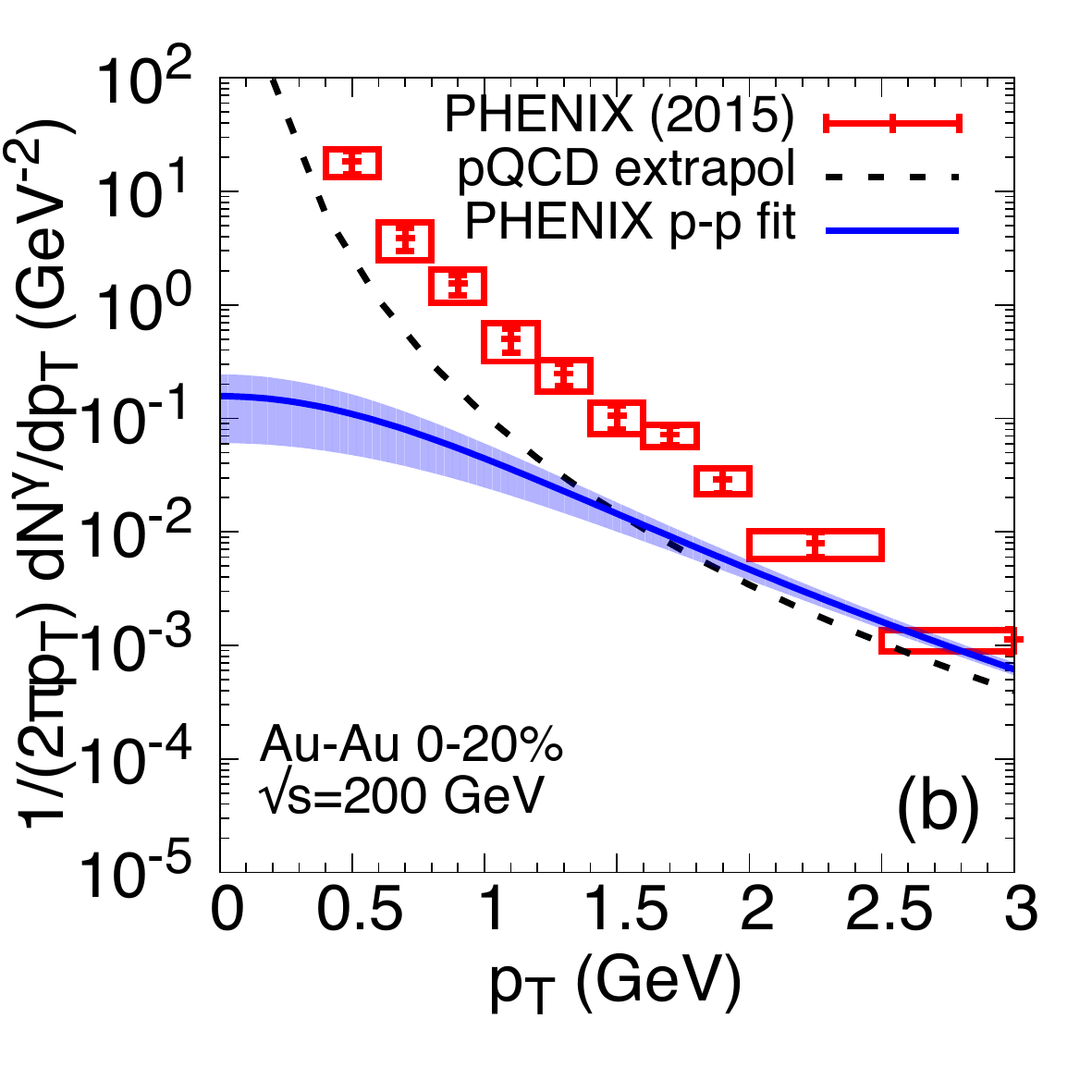}
				\hfill
				\includegraphics[width=0.32\textwidth]{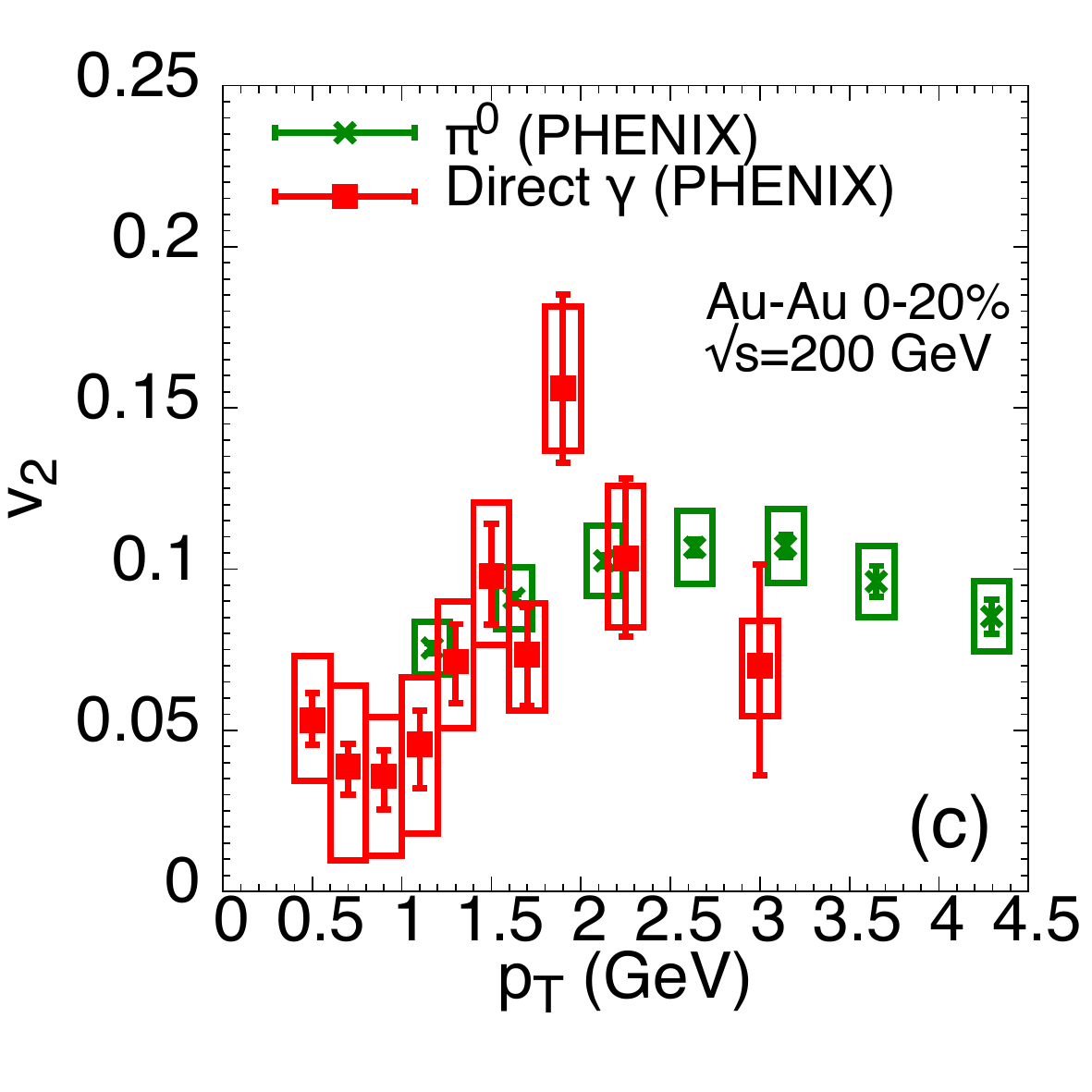}
		 \caption{(a) Hadronic and charm quark dilepton decays compared to the dilepton invariant mass spectra from the STAR collaboration, for minimum bias Au+Au collisions at $\sqrt{s_{NN}}=200$~GeV~\cite{Adamczyk:2015lme} (b) Direct photon spectra in 0-20\% centrality Au+Au collisions at $\sqrt{s_{NN}}=200$~GeV~\cite{Adare:2014fwh}, compared to the prompt photon background as predicted by a fit~\cite{Adare:2014fwh} from the PHENIX collaboration to direct photon measurements in p-p collisions and to a perturbative QCD calculation~\cite{Paquet:2015lta} (c) Direct photon $v_2$ \cite{Adare:2015lcd} in the same collision system as (b), compared to the $v_2$ of $\pi^0$~\cite{Adare:2011zr}.}
				\label{fig:coldphotons}
\end{figure}

One of the large ``cold'' sources of dileptons is hadronic decays 
(e.g. $\pi^0\to \gamma e^+ e^- $, $\eta \to \gamma e^+ e^- $,  $\omega \to e^+ e^-$)
, with different hadrons contributing in different windows of dilepton invariant mass $M_{l^+l^-}$. The contribution to the dilepton invariant mass spectra of decays channels with three-body final states (e.g. $\pi^0\to \gamma e^+ e^- $) dies off when $M_{l^+l^-}$ exceeds the mass difference between the decaying hadron (e.g. $\pi^0$) and the daughter particle accompanying the lepton pair (e.g. $\gamma$).
Hadronic decays with only $ e^+ e^- $ in the final state (e.g. $\omega \to e^+ e^-$) rather produce a peak in the spectra at an invariant mass close to the decaying hadron's mass.
These different hadronic decay channels, combined with branching fractions that vary considerably from channel to channel, produce a complicated but fairly well-understood background, which partially hide the thermal dilepton signal. 
Dileptons from charm quark decays are also a significant background over a wide invariant mass range ($\sim 0-4$~GeV). 
Dileptons from the hard initial nucleon-nucleon scattering --- Drell-Yan dileptons --- are typically a small contribution at low values of $M_{l^+l^-}$. 
The hadronic and charm dilepton backgrounds are shown in Figure~\ref{fig:coldphotons}(a) for Au+Au collisions at $\sqrt{s_{NN}}=200$~GeV with the STAR detector's acceptance cuts~\cite{Adamczyk:2015lme}. These backgrounds are compared with the total (inclusive) dilepton measurement from STAR~\cite{Adamczyk:2015lme}, and show an excess of dileptons in the $0.3-0.6$~GeV and $0.8-1.0$~GeV invariant mass regions. This excess is not seen in proton-proton collisions~\cite{Adare:2008ac,Adamczyk:2012yb}, and is interpreted as a signal from the deconfined plasma produced in heavy-ion collisions. Since dilepton decays from certain hadrons (in particular the $\rho$ meson) and from charm quarks are affected by the presence of the hot plasma in heavy-ion collisions, the difference between dilepton measurements in Au+Au collisions and the ``cold background'' shown on Fig.~\ref{fig:coldphotons}(a) is understood to be a combination of both a thermal dilepton signal and a modification of cold dilepton sources (see e.g. Ref.~\cite{Rapp:2013nxa} for a recent overview).

The dilepton spectra can also be investigated as a function of the transverse momentum $p_T$ of the lepton pair. Since different sources of dileptons dominate the signal in different ranges of transverse momentum $p_T$ and invariant mass  $M_{l^+l^-}$, it is possible to construct a variety of observables more and less sensitive to thermal dileptons and the different backgrounds. One can look at the invariant mass spectra with various cuts in $p_T$, or at the $p_T$ spectra with various cuts in  $M_{l^+l^-}$ (see e.g. Ref.~\cite{Adamczyk:2015lme}). Beyond the spectra, the momentum anisotropy of dileptons is also of interest, although this is a difficult measurement and only limited data are available at the moment~\cite{Adamczyk:2014lpa}.

\paragraph{Photons}

For photons, hadronic decays are the largest cold source of radiation at low transverse momentum ($p_T^\gamma \sim 0-4$~GeV), with $\pi^0 \to \gamma \gamma$ dominating the signal, followed by $\eta \to \gamma \gamma$. The main difference with dileptons is that the simpler kinematic of photon decays makes it possible to subtract statistically the contribution of hadronic decays from photon measurements. The resulting observable, called ``direct photons'', provides a better access to the thermal photon signal than photon measurements that include hadronic decays.

The single significant ``cold'' source of photons present in direct photon measurements are prompt photons
. These photons are produced in hard initial nucleon scattering, at the very initial impact of the colliding ions, and are known to scale with the number of binary collisions (discussed above).
Prompt photons can be produced either when a hard parton from each nucleon produces a final state photon directly (e.g. $q+\bar{q}\to g + \gamma $), or when a hard final state parton is produced (e.g. $q+q\to q + q $) and later radiates a photon (so-called \emph{fragmentation photons}). These processes have been computed perturbatively at next-to-leading order in the strong coupling constant $\alpha_s$~\cite{Aurenche:1987fs,Aversa:1988vb}. In proton-proton collisions, these perturbative QCD calculations of prompt photons have been found to be in very good agreement with direct photon measurements
at both RHIC and the LHC~\cite{Adare:2012yt,Chatrchyan:2012vq,Aad:2013zba}.
This dominance of prompt photons in \emph{proton-proton} collisions implies that their contribution in \emph{heavy-ion} collisions can be obtained using either direct photons measurements from p-p collisions or perturbative QCD calculations of prompt photons. Both methods give essentially the same result at high $p_T$ ($p_T^\gamma \gtrsim 5$~GeV), and prompt photons describe very well the direct photon signal at large $p_T^\gamma$ in heavy-ion collisions~\cite{Chatrchyan:2012vq,Adam:2015lda,STAR:2016use}.

At low $p_T^\gamma$, the contribution of prompt photons to heavy-ion collisions is a more difficult question. Direct photon measurements in p-p collisions are not yet very precise at low $p_T^\gamma$ ($p_T^\gamma \lesssim 1.5$~GeV). On the other hand, it is unclear how reliable are perturbative QCD calculations for small photon energies. Moreover, prompt photon production is expected to be different in heavy-ion collisions than it is in p-p collisions, because final state parton interactions with the deconfined plasma modifies fragmentation photon production~\cite{Fries:2002kt,Turbide:2005fk,Turbide:2007mi,Arleo:2011gc,Zakharov:2004bi}. 
This complication does not appear at high $p_T^\gamma$, where the fragmentation photon contribution is small.

At the moment, 
the contribution of prompt photons to heavy-ion collisions is typically estimated with binary-nucleon-collision-scaled direct photon measurements from proton-proton collisions, or with perturbative QCD calculations without parton energy loss. Such estimates of prompt photons are shown in Figure~\ref{fig:coldphotons}(b), and are compared with direct photon measurements in Au+Au collisions at RHIC. One can see that while the two estimates of prompt photons in heavy-ion collisions differ\footnote{The large difference between the two prompt photon estimates at low $p_T^\gamma$ is straightforward to understand. The lower estimate is from a fit~\cite{Adare:2014fwh} by the PHENIX collaboration to direct photons measurements in p-p collisions. This fit is still poorly constrained by measurements below $p_T^\gamma\sim 1.5$~GeV. Moreover, the fit and its uncertainty band rely on the assumption that prompt photon production saturates for $p_T^\gamma$ lower than $1-2$~GeV. On the other hand, the prompt photon estimate based on perturbative QCD assumes that the power-law dependence of the spectra observed at high $p_T^\gamma$ continues at low $p_T^\gamma$~\cite{Paquet:2015lta}. This last scenario may considerably overestimates low $p_T^\gamma$ prompt photon production. New low $p_T^\gamma$ measurements of direct photons in p-p collisions, together with calculations of prompt photons that include the effect of the hot deconfined plasma on fragmentation photons, will help clarify the situation.
}
below $p_T^\gamma \sim 1.5$~GeV, they are both considerably lower than the direct photon measurements. This means that direct photon measurements in heavy-ion collisions are showing a considerable excess above prompt photons, an excess generally attributed to thermal photons.

Measurements are also available for the momentum anisotropy $v_n$ of direct photons~\cite{Adare:2015lcd,Lohner:2012ct}, obtained as for the spectra by a statistical subtraction of hadronic decay photons. The measured direct photon $v_2$ is found to be of the same order as the pion $v_2$, as shown on Fig.~\ref{fig:coldphotons}(c) for Au+Au collisions at RHIC. This is a very large $v_2$, considering that prompt photons are expected to have a momentum anisotropy around an order of magnitude lower than pions~\cite{Turbide:2007mi}. This is again seen as a strong signal of thermal photon production.
A closer look at the momentum anisotropy development of thermal photons is provided in the next section.

One should note that other sources of electromagnetic probes in heavy-ion collisions have been suggested. A recent overview can be found in Refs.~\cite{Shen:2015nto,Shen:2016odt}. The importance of such sources is still under investigation. 
The rest of this work focuses on thermal photons.

\section{Thermal emission}

Thermal electromagnetic probes are those emitted by the deconfined plasma during its phase of hydrodynamical expansion, while the plasma is close to local thermal equilibrium. Hydrodynamical models of heavy-ion collisions provide a description of the deconfined nuclear plasma, from hyperbolic time $\tau \sim 0.1-1$~fm to $\sim 10$~fm, in terms of macroscopic variables: the local temperature $T(X)$ of the plasma, the flow velocity profile $u^\mu(X)$ and the extent of the plasma's deviation from equilibrium, as quantified by the shear stress tensor $\pi^{\mu\nu}(X)$ and bulk pressure $\Pi(X)$. Given local values for $T$, $u^\mu$, $\pi^{\mu\nu}$ and $\Pi$, the production of thermal photons and dileptons in a small spacetime volume $\Delta^4 X$ is equal to the thermal electromagnetic emission rate $d^4 \Gamma_{\gamma/l^+l^-}(K^\mu,u^\mu,T,\pi^{\mu\nu},\Pi)/d^4 k$ multiplied by $\Delta^4 X$. The total thermal electromagnetic probe signal is obtained by integrating the emission rate over the spacetime profile provided by hydrodynamics:
\begin{equation}
 \frac{d^4 N_{\gamma/l^+l^-}}{d^4 k}=\int d^4X \frac{d^4 \Gamma_{\gamma/l^+l^-}}{d^4 k}(K^\mu,u^\mu(X),T(X),\pi^{\mu\nu}(X),\Pi(X))
\end{equation}
where the integral over $X$ is usually limited to the spacetime volume that has not reached kinetic freeze-out (which usually means the region with $T>T_{FO}$ when the kinetic freeze-out criteria for the hydrodynamical model is the temperature $T_{FO}$). The thermal or quasi-thermal  photon/dilepton emission rate $d^4 \Gamma_{\gamma/l^+l^-}/d^4 k$ is constrained theoretically from different models of hadronic and partonic degrees of freedom. Recent summaries of the thermal emission rates, including discussions of the effects of shear and bulk viscosities, can be found in e.g. Refs.~\cite{Shen:2015nto,Shen:2016odt,Vujanovic:2016anq,Paquet:2015lta}. Uncertainties remain in $d^4 \Gamma_{\gamma/l^+l^-}/d^4 k$, especially in and slightly above the cross-over temperature region ($T\sim 150-300$~MeV), and a better understanding of the thermal electromagnetic rates is still an active field of research.

Photons and dileptons inherit a small momentum anisotropy from the thermal emission rate $d^4 \Gamma_{\gamma/l^+l^-}/d^4 k$, which is not isotropic when non-equilibrium (viscous) corrections are taken into account. However, as for soft hadrons, the momentum anisotropy of thermal electromagnetic probes originates overwhelmingly from a different mechanism: the asymmetry of the plasma's flow velocity profile $u^\mu(X)$. Initially ($\tau \sim 0.1-1$~fm ), flow velocities are thought to be small and similar in size along the short and long axes of the almond-shape energy deposition in heavy-ion collisions. Nevertheless, after $\sim 10$~fm of hydrodynamic evolution, the flow is much stronger along the short axis of the energy deposition. This flow anisotropy gives thermal photons and dileptons emitted at late times a $v_n$ similar to hadrons, while electromagnetic probes emitted at early times have much smaller momentum anisotropies. To obtain the large $v_2$ of direct photons seen in Fig.~\ref{fig:coldphotons}(c) from a hydrodynamical calculation of thermal photons, one needs a rapid development and saturation of the flow velocity anisotropy in heavy-ion collisions, as well as a significant production of thermal photons at late times. Since prompt photons are also present and have a small momentum anisotropy, thermal photons must also outshine them in order for a large $v_2$ to be seen in the sum of their contributions.

At the moment, the combination of a prompt photon background with hydrodynamical calculations of thermal photons generally underestimates measurements of the direct photon spectra and $v_2$ at RHIC and the LHC~\cite{Shen:2015nto,Shen:2016odt}. On the other hand, these calculations describe many \emph{qualitative} features observed in measurements, such as an exponential  $p_T^\gamma$-dependence of the direct photon spectra at low $p_T^\gamma$, and a large thermal photon $v_2$ that agrees well with measurements before prompt photons are added. In this sense, targeted improvements to current hydrodynamic calculations, such as a better description of the early stage of the plasma, possible changes in the thermal emission rates in the cross-over region and a revisiting of low $p_T^\gamma$ prompt photon production could prove sufficient to resolve the tension observed with measurements. Nevertheless, the presence of additional sources of photons still cannot be ruled out and investigations in that direction are on-going.

While thermal photons have been studied over the past decade with increasingly sophisticated hydrodynamical models, hydrodynamical studies of thermal dileptons have been fewer, possibly due to the limited availability of measurements. A recent hydrodynamic calculation~\cite{Vujanovic:2013jpa} found general good agreement with the low-$M_{e^+e^-}$ invariant mass spectra measured by the STAR collaboration. This result is in line with earlier calculations made with a simpler description of the plasma evolution~\cite{Rapp:2013nxa}.  

The reasons behind the different levels of agreement with the available dilepton and photon measurements of hydrodynamic calculations are still under investigation.
An important step will be to evaluate both thermal photons and dileptons from the same modern hydrodynamic simulation, with similarly constrained thermal emission rates, to provide for a more consistent comparison with measurements. It must also be emphasised that the dilepton invariant mass spectra is integrated in $p_T^{e^+e^-}$, which greatly reduces its sensitivity to the flow velocity profile of the plasma. In this sense, the photon and dilepton measurements currently available are sensitive to different features of the deconfined plasma.

As more systematic comparisons of hydrodynamical models with electromagnetic and hadronic observables are made, it should become clearer if agreement with measurements can be improved by modifying the spacetime description of the plasma evolution, or by including new sources of radiation, or even simply by a better understanding of the thermal emission rates.


\ack I thank Charles Gale, Chun Shen and Gojko Vujanovic, along with the organizers and participants of the Hot Quarks 2016 workshop, for interesting discussions and feedback. This work was supported by the U.S. D.O.E. Office of Science, under Award No. DE-FG02-88ER40388.

\section*{References}
\bibliography{biblio}

\end{document}